# ARTICLE

# Minimum Surfactant Concentration Required for Inducing Self-shaping of Oil Droplets and Competitive Adsorption Effects

Jiale Feng[a, e], Zhulieta Valkova[b], E Emily Lin[d], Ehsan Nourafkan[d], Tiesheng Wang[a,c], Slavka Tcholakova[b*], Radomir Slavchov[d], Stoyan K. Smoukov[a, b, d*]

Surfactant choice is key in starting the phenomena of artificial morphogenesis, the bottom-up growth of geometric particles from cooled emulsion droplets, as well as the bottom-up self-assembly of rechargeable microswimmer robots from similar droplets. The choice of surfactant is crucial for the formation of a plastic phase at the oil-water interface, for the kinetics, and for the onset temperature of these processes. But further details are needed to control these processes for bottom-up manufacturing and understand their molecular mechanisms. Still unknown are the minimum concentration of the surfactant necessary to induce the processes, or competing effects in a mixture of surfactants when only one is capable of inducing shapes. Here we systematically study the effect of surfactant nature and concentration on the shape-inducing behaviour of hexadecane-in-water emulsions with both cationic (CTAB) and non-ionic (Tween, Brij) surfactants over up to five orders of magnitude of concentration. The minimum effective concentration is found approximately equal to the critical micelle concentration (CMC), or the solubility limit below the Krafft point of the surfactant. However, the emulsions show low stability at the vicinity of CMC. In a mixed surfactant experiment (Tween 60 and Tween 20), where only one (Tween 60) can induce shapes we elucidate the role of competition at the interface during mixed surfactant adsorption by varying the composition. We find that a lower bound of ~ 75% surface coverage of the shape-inducing surfactant with C14 or longer chain length is necessary for self-shaping to occur. The resulting technique produces a clear visual readout of otherwise difficult to investigate molecular events. These basic requirements (minimum concentration and % surface coverage to induce oil self-shaping) and the related experimental techniques are expected to guide academic and industrial scientists to formulations with complex surfactant mixtures and behaviour.

## 1. Introduction

Bottom-up fabrication promises to replace energy- and material-wasteful top-down processes such as lithography with sustainable, scalable processes, though the types of shapes and control over them have been limited.[1–3] The recent discovery of artificial morphogenesis can lead to scalable manufacturing of particles of a large variety of regular geometric shapes, simply by cooling down emulsion droplets.[4–6] These include polymer particles[7,8] in the shapes of polyhedra, hexagons, rhomboids, triangles, rods, and fibres. The mild conditions and lack of subtractive steps promise a sustainable and efficient process for fabrication by bottom-up growth. We have control of the size of these shapes over orders of magnitude, from as small as 50 nm to over 1 mm.[7] On a fundamental level, the process is started by the formation of a few layers of plastic crystal (rotator phase) in the oil just near the water interface, which exert stress opposing the interfacial tension and generate anisotropic shapes in an established sequence.[9,10] The formation of the plastic crystal phase depends critically on the molecular properties of the surfactant used. Alternative mechanisms have been discussed in the literature. Garcia-Aguilar et al. proposed an elastic surface model;[11] we compared this hypothesis to ours.[12]

We have described the length of the surfactant tail needed to induce the transition, and have classified the behaviour of many classes of surfactants according to their ability to generate the shapes at a constant surfactant concentration.[5] We have successfully modelled the shape evolution sequence as a competition between the fundamental thermodynamic driving forces for the formation of the plastic crystal phase and the opposing increase of surface energy due to interfacial tension.[9,10] We have recently also shown that the phenomenon is quite general, with many classes of oils showing this behaviour,[5] and also that even oils that do not by themselves undergo such changes can be induced to do so in mixtures with other oils.[13] The phenomenon was demonstrated also for

[a.] *Active and Intelligent Materials Lab, Department of Materials Science & Metallurgy, University of Cambridge, 27 Charles Babbage Road, Cambridge CB3 OFS, UK*
[b.] *Department of Chemical and Pharmaceutical Engineering, Faculty of Chemistry and Pharmacy, Sofia University, 1 James Bourchier Ave., 1164 Sofia, Bulgaria*
[c.] *School of Mechanical Engineering, Shanghai Jiao Tong University, Shanghai, 200240, China*
[d.] *School of Engineering and Materials Science, Queen Mary University of London, Mile End Road, London E1 4NS, UK*
[e.] *Cavendish Laboratory, Department of Physics, University of Cambridge, JJ Thomson Avenue, Cambridge CB3 0HE, UK*

\* Corresponding authors: Stoyan K. Smoukov, email: s.smoukov@qmul.ac.uk; Slavka Tcholakova, email: sc@lcpe.uni-sofia.bg

Electronic Supplementary Information (ESI) available: [details of any supplementary information available should be included here]. See DOI: 10.1039/x0xx00000x

water-in-oil emulsions and at flat water|oil interfaces.[14,15] We have only recently developed the tools of molecular dynamics to simulate liquid-solid and solid-solid crystal-rotator phase transitions that may allow soon to reveal the molecular details of this process.[16] Still, many details of the mechanism remain unknown[13] and behaviour is hard to predict. For example, small changes in surfactant structure or difference between surfactant tail and oil, have caused the emergence of self-assembled swimmers[14].

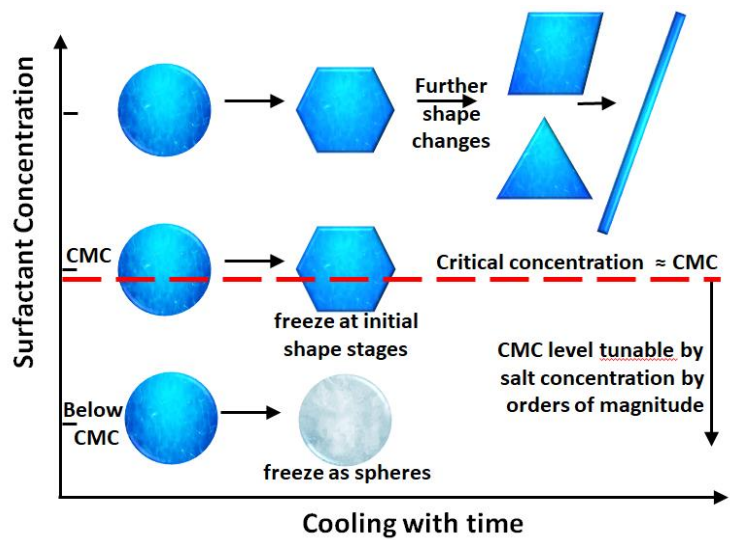

Figure 1: Schematic representation of the experiment. Hexadecane emulsions were prepared in aqueous surfactant solution. Upon cooling, when below the CMC (or below solubility, in the case of CTAB), droplets freeze as spheres. At or near CMC they exhibit shape changes on cooling. CMC and therefore surfactant packing on the surface is important, as the same absolute concentration of surfactant that could not induce shape changes, does in the presence of salt where CMC is lower. At high concentrations (> several times CMC), droplets deform all the way along the shape-change sequence we have established for the process of artificial morphogenesis.[4]

One of the open questions is about the minimum concentration of surfactant and the type of packing at the surface that is necessary for the plastic crystal templating to occur.[17,18] There are even more complicated questions of how mixtures of surfactants would affect the packing on the surface and rotator phase formation.

In this report, we conduct systematic experiments into the minimum surfactant concentration necessary to induce the self-shaping. Two classes of surfactants are studied: non-ionic families Tween (Tween 40 and Tween 60) and Brij (Brij S10, S20, C10 and Brij58) and cationic hexadecyltrimethylammonium bromide ($C_{16}$TAB). For all the surfactants we show the minimum concentration necessary to induce the transformation, with or without salt in the solution, is close to the critical micellar concentration (CMC) (Figure 1). Near the CMC the droplets can only transform to the initial shapes of the prescribed shape evolution sequence. At higher concentrations, they transform to the end of the sequence (Figure 1). We also perform competitive adsorption experiments with binary mixtures of surfactants, only one of which can induce self-shaping.

## 2. Experimental Section

### 2.1. Materials

As dispersed oily phase in the studied emulsions, we used the linear alkane hexadecane ($C_{16}$), obtained from Sigma-Aldrich with purity ≥ 99%, melting temperature $T_m$ = 18 °C, used as received, without further purification.

To stabilize the emulsions, we used a variety of nonionic and ionic surfactants. The nonionic surfactants were from two main groups Tween and Brij surfactants. Polyoxyethylene sorbitan monoalkylates (Tween surfactants, $C_n SorbEO_{20}$) differed in the length of their alkyl chain, $n$, varied between 12 and 18 (Tween 20, Tween 40, and Tween 60); the number of oxyethylene units, $m$, is ~20. The polyoxyethylene alkyl ethers (Brij surfactants, $C_n EO_m$) were with $n$ = 16 or 18 carbon atoms and $m$ = 10 or 20 oxyethylene units. We used a cationic surfactant with hexadecyl tail – $C_{16}$TAB. Detailed information for the purity, HLB values of the nonionic surfactants and surfactant producers is presented in the Supporting information, Table S1. All surfactants were used as received. For experiments in the presence of salt we used NaBr, a product of Sigma with purity ≥ 99.5%.

As a fluorescent dye, we used the oil soluble Solvent Green 5 (diisobutyl perylenedicarboxylate), produced by TCI Chemicals. The addition of the fluorescent dye brings two main benefits: (1) helps to observe some finer structures and eliminates the stains on the background, (2) acts as an indicator in the membrane emulsification process (see Methods below), distinguishing droplets from small air bubbles. The concentration of fluorescent dye in $C_{16}$ drops is 0.1 wt. %.

All aqueous solutions were prepared with deionized water, which was purified by an Elix 3 module (Millipore, USA).

### 2.2. Methods

**2.2.1.    Preparation of the initial emulsions.**

The initial alkane-in-water emulsions were prepared by three different experimental procedures. In the first we prepare each emulsion at the specified concentration, though the emulsion stability is poorer. In the second we prepare a stable emulsion at high concentration and dilute it as necessary. In the third procedure, the double syringe technique was applied for the preparation of emulsions.

*Experimental procedure 1:  Preparation of polydisperse emulsion droplets in target solution concentration*

First, we prepared an aqueous surfactant solution with fixed surfactant concentration. Afterwards, we added the alkane phase into the aqueous phase. The alkane-in-water emulsions by the first experimental procedure were prepared by homogenizing the disperse phase on a magnetic stirrer for 15 minutes and after that, a specimen of the studied emulsion was placed in a capillary for the optical observations. In this procedure, the experiments with the emulsions were performed without further dilution of the droplets. The droplet size range is 2-80 μm.

*Experimental procedure 2: Preparation of monodisperse emulsion droplets in stock solution to be diluted later*

The second set of emulsions were prepared by a laboratory Microkit membrane emulsification module from Shirasu Porous Glass Technology (SPG, Miyazaki, Japan). The oil phase was passed through tubular glass membranes with outer diameter of 10 mm and working area of approximately 3 cm$^2$. This method allows production of drops with relatively narrow drop size distribution.[19–22] Membranes with mean pore diameter of 5 μm were used for preparing initial emulsions resulting in drop diameters of around 15-20 μm. There are several advantages of the membrane emulsification method compared with classical emulsification: (1) narrow size distribution of droplets, (2) little heat generated due to no high pressure/shear stress required, (3) the energy input is quite low while the energy efficiency is high.

In this experimental procedure, we kept the surfactant concentration sufficiently high during preparation of the monodisperse emulsion drops, i.e. 1.5 wt. % for the non-ionic surfactants and 0.5 wt. % for the cationic surfactant. After that, we diluted the alkane-in-oil emulsion with distilled water until the desired surfactant concentration was reached. In addition, we stirred gently the sample for 15 minutes to allow for surfactant redistribution between the solution and droplets.

The main difference between these two experimental procedures is the initial surfactant concentration in which the $C_{16}$ droplets are emulsified.

*Experimental procedure 3: Preparation of monodisperse emulsion droplets using double syringe technique*

Emulsions stabilized by Brij surfactants were prepared by making an aqueous surfactant solution at the final concentration, adding the required amount of oil, and passing the mixture between two syringes, connected by a double-sided Luer-Lock 18 gauge micro-emulsifying needle (stainless steel, 18G * 2-7/8'' with reinforcing bar, Cadence Science, Inc.). The syringe sizes were 5 cm$^3$ plastic syringes with Luer Lock connectors, and care was taken to remove all air from the system, including the needle, before emulsification. The mixture was passed at high speed through the needle between the two syringes. 15-50 passes were made, depending on the surfactant concentration and the resulting emulsion was used as obtained, without further dilution.

**2.2.2.    Optical observations of the drop shape transformations.**

For microscope observations, a specimen of the studied emulsion was placed in a capillary with rectangular cross section: 50 mm length, 1 mm width, and 0.1 mm height. The capillary was enclosed within a custom-made metal cooling chamber, with optical windows for microscope observation. The chamber temperature was controlled by cryo-thermostat (JULABO CF30, Cryo-Compact Circulator) and measured close to the emulsion location, using a calibrated thermo-couple probe with an accuracy of ± 0.2°C. The initial temperature for cooling droplets was 20 °C. The thermo-probe was inserted in one of the orifices of the aluminum thermostatic chamber and mounted in the position where a capillary with the emulsion sample would be normally placed for microscope observations. In the neighboring orifice, the actual capillary with the emulsion sample was placed. The correct measurement of the temperature was ensured by calibrating the thermo-couple with a precise mercury thermometer in the respective range of temperatures measured. Furthermore, we always observe melting of the frozen particles at temperatures within ± 0.2°C of the melting temperature of the bulk oil, reported in the literature.

We also prepared another set of experiments in which the temperature of the sample is controlled by a Linkam cooling system for regulating temperature and cooling rate. The temperature limits and cooling rate can be set up on the control panel, with temperature ranging from -25℃ to 120℃ and cooling rate ranging from 0.01 K/min to 20 K/min.

The optical observations were performed with AxioImager.M2m (Zeiss, Germany) and Olympus BX51 microscopes. Long-focus objectives 20x, 40x, 50x, 60x and 100x were used. For the experimental observations performed in transmitted, cross-polarized white light, we used λ (compensator) plate, situated after the sample and before the analyzer, at 45° with respect to both the analyzer and the polarizer. Under these conditions the liquid background and the fluid objects have typical magenta color, whereas the birefringent areas appear brighter and may have intense colors.[23,24] For the experiments with emulsion droplets containing a fluorescent dye a reflective light was used.

**2.2.3.    Interfacial tension (IFT) measurement.**

The pendant drop tests were performed using Kruss Tensiometer (DSA 100) to measure the IFT and estimate the CMC of the surfactants. A calibration test was performed by surface tension measurement of deionized water before the start of each set of analysis. A glass cuvette made of optical glass and a hooked needle (0.74 mm diameter) were used to measure IFT between hexadecane/aqueous phases.

## 3. Results and Discussion

**Surfactant concentration effects on self-shaping**

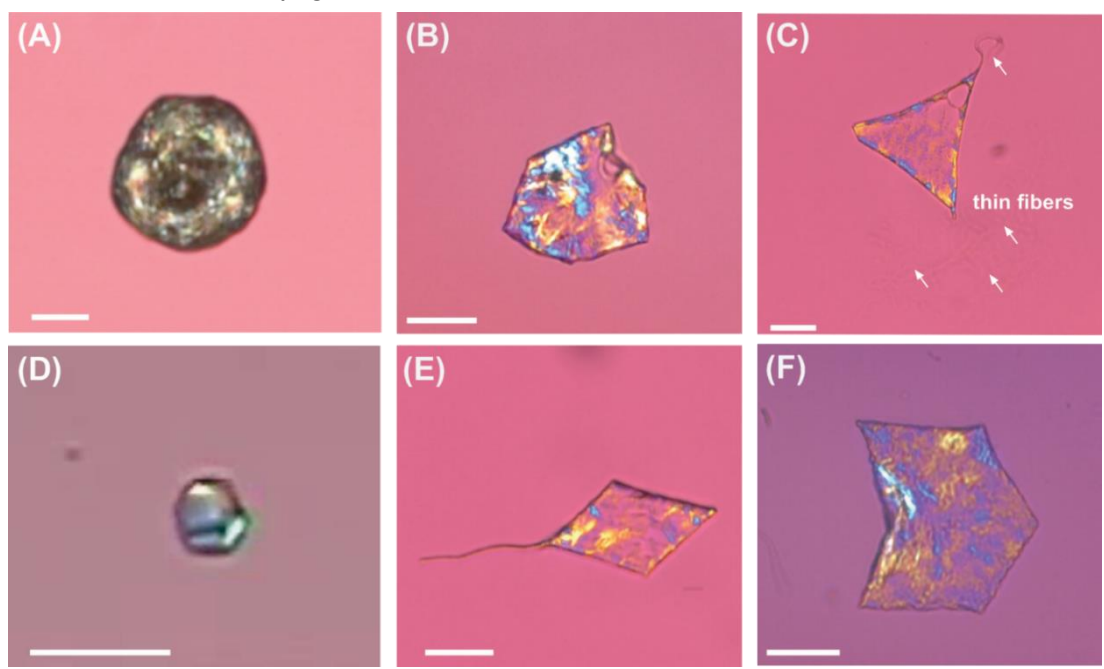

Figure 2: Microscopic pictures of frozen $C_{16}$ particles stabilized by $C_{16}$TAB with different surfactant concentrations. **(A)-(C) in the absence of salt, CMC ≈ 0.98 mM**: **(A)** 0.75 mM, **(B)** 1.0 mM and **(C)** 10 mM of $C_{16}$TAB. By increasing the surfactant concentration, we increase the probability of formation of different flat shapes. At 0.75 mM $C_{16}$TAB, droplets freeze into a spherical shape, while at 10 mM $C_{16}$TAB we observe formation of flat shapes, which are extruding long thin fibres. Cooling rate 0.44 K min$^{-1}$. Scale bar 20 μm. **(D)-(F) in the presence of salt (100 mM NaBr)**, **CMC ≈ 0.04 mM**: **(D)** 0.05 mM, **(E)** 0.4 mM and **(F)** 0.5 mM $C_{16}$TAB. Cooling rate 0.44 K min$^{-1}$. Scale bar 20 μm.

We present results obtained with $C_{16}$ emulsion droplets stabilized by different surfactants, with different surfactant concentrations, and in the presence and absence of salt. In **Section 3.1** we describe the surfactant concentration effects for cationic surfactant on the self-shaping phenomena. In **Section 3.2** we describe these effects for non-ionic surfactants. The competitive adsorption effects on self-shaping in mixtures of nonionic surfactants are presented in **Section 3.3**.

### 3.1. Concentration effects for cationic surfactant

We set out to find the minimum surfactant concentration for the cationic surfactant cetyltrimethylammonium bromide ($C_{16}$TAB that induces the self-shaping behaviour of the hexadecane ($C_{16}$) droplets. For ionic surfactants, both solubility limit and CMC are greatly affected by concentration of the added salt.[25,26] Therefore, we determined the minimum required surfactant concentration for self-shaping of the droplets both in the absence and in presence of salt. We performed one set of experiments with surfactant in pure water and another set with the same surfactant in the presence of 100 mM NaBr.

$C_{16}$TAB is below its Krafft temperature in both cases. Thermal hysteresis is, however, very common for this surfactant – it can exist as metastable micelles for days even at concentrations 20 times higher than the solubility limit, 10 °C below the Krafft temperature.[27] To determine the CMC/solubility value in each case, we measured the interfacial tension of the hexadecane/water interface as a function of the surfactant concentration. According to the data from the literature, the CMC of $C_{16}$TAB at room temperatures in the absence of salt is around 0.98 mM and in presence of 100 mM NaBr is ~ 0.04 mM.[28] The solubility of the surfactant is difficult to measure, but at the Krafft temperature (26.6 °C in the absence of salt),[27] it is equal to CMC, therefore, the solubility product is $K_s = [C_{16}TA^+][Br^-] = (0.98 \text{ mM})^2$. At $[Br^-] = 100$ mM, this corresponds to solubility $[C_{16}TA^+] = K_s/[Br^-] = 0.01$ mM. The results that we obtained from the breakpoint of the measured interfacial tension vs. $[C_{16}TA^+]$ are in better agreement with the CMC values: 0.99 mM for the system without salt and 0.03 mM in the presence of 100 mM NaBr.

Below the breakpoint (be it solubility limit or CMC), the $C_{16}$ emulsion drops do not transform their shape at all and freeze into spherical shape at temperatures between 8-10 °C, see Figure 2A for $[C_{16}TAB] = 0.75$ mM, no salt. Similar results were obtained with 0.8 and 0.9 mM $C_{16}$TAB. At 1.0 mM, we are able to observe drop shape transformations, see Figure 2B, where emulsion drops with $d_{ini} \approx 10$-25 μm reached the initial 3D shapes of the transformation, and only transformed to the next flat platelet stage if we decreased the cooling rate to ≈ 0.2 K min$^{-1}$. When we increase the surfactant concentration 10 times (10 mM $C_{16}$TAB), we are able to observe formation of flat platelets which are extruding long fibers, see Figure 2C. Droplets with smaller initial drop size diameter (< 10 μm) are able to reach the final stages of the evolutionary scheme forming ellipsoidal droplets connected with long thin fibres even at 1 mM concentrations.

Similarly, in the case of salt added, we first observed deformations around the interfacial tension breakpoint, though in this case at much lower surfactant concentrations, because the addition of salt decreases both CMC and solubility of the surfactant. At surfactant concentration below 0.04 mM no shape changes are observed, while in the range between 0.04 mM and 0.2 mM, we are able to observe deformations only in the small droplets ($d_{ini} \leq 7$ μm), see Figure 2D. At higher concentrations ($\geq$ 0.3 mM, prepared by heating and cooling back to room temperature), the bigger droplets also start to transform, see Figure 2E, F. The obtained frozen particles are identical to those in the absence of NaBr. For example, close to CMC/solubility limit, the drops are only able to transform their shape to the initial stages of the evolutionary scheme, forming regular polyhedral and other faceted 3D shapes. At higher surfactant concentration ($\geq$ 0.4 mM $C_{16}TAB$, Figure 2B), the drops are able to transform their shape into flat platelets (rhomboid, triangular and hexagonal prisms) extruding long thin fibres. With this series of experiments we showed that by the addition of an electrolyte to the water phase, we are able to reduce the concentration of the ionic surfactant necessary to produce shapes from 1 mM to 0.04 mM (~ 25 fold).

Tweens were chosen for these experiments as they are edible, biocompatible and are the most common commercial non-ionic class of surfactants. They are polysorbate molecules each containing a hydrophilic head group of oligo (ethylene glycol) chains and a hydrophobic tail of fatty acid ester moiety (Table S1).[29] They are used as emulsifiers and solubilizers from pharmaceuticals to food products, including bread, cake mixes, salad dressings, shortening oil and chocolate. In addition, they are often used in cosmetics to solubilize essential oils into water-based products. Both Tween 40 (approx. 90% $C_{16}SorbEO_{20}$ + 10% $C_{18}SorbEO_{20}$) and Tween 60 (approx. 50% $C_{16}SorbEO_{20}$ + 50% $C_{18}SorbEO_{20}$) have been shown previously to induce self-shaping in hexadecane emulsion droplets.

In our general classification of shape-inducing surfactants[5] these two fall into **Group B**. For this group, shape transformations start around the oil melting temperature, i.e. $T_d \approx T_m$. They are able to form very strong (thick) interfacial layers and typically form hexagonal and other platelets, as well as relatively stiff rods. These include formation of perforated fluid platelets that maintain their outer shape integrity after the thin liquid film in their middle spontaneously breaks. All previously reported results for polysorbate surfactants are

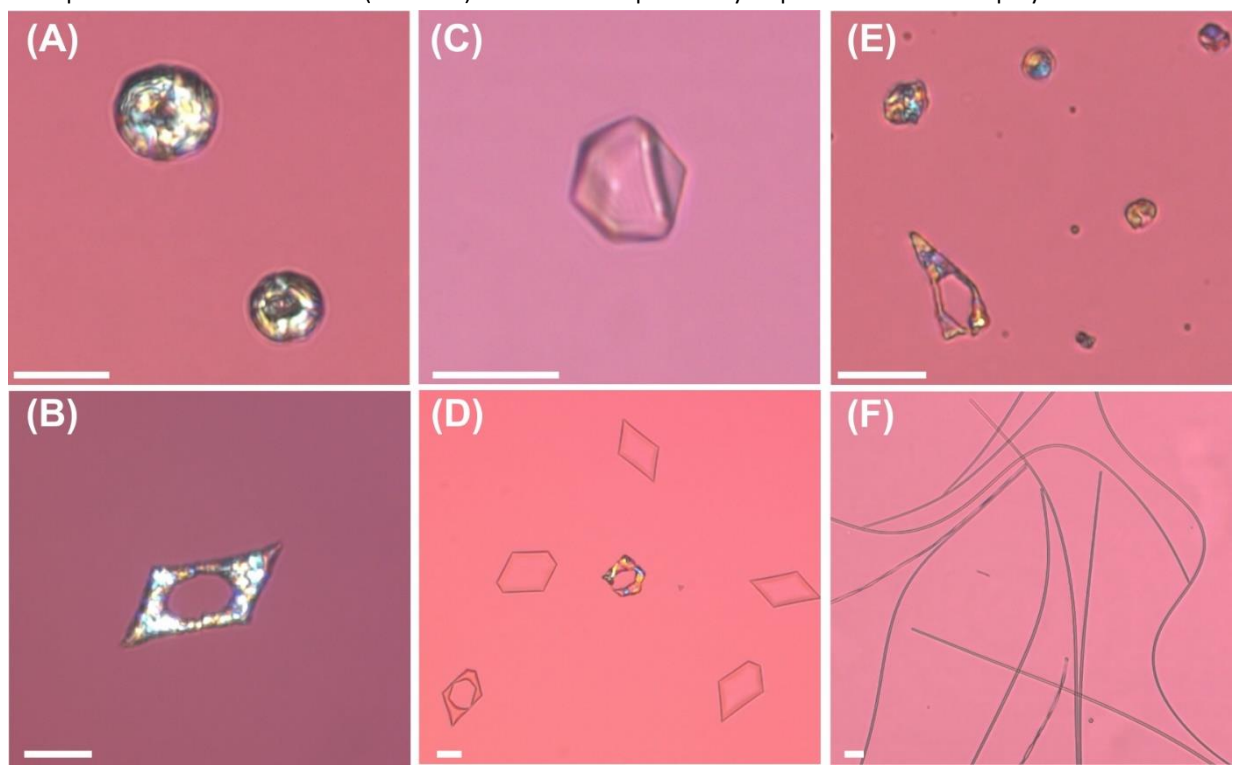

Figure 3: **Drop shape transformations for emulsions prepared** from $C_{16}$ oil drops stabilized by Tween 40 ($C_{16}SorbEO_{20}$) **by two different procedures**. Images on the **first row** are obtained with emulsion droplets prepared by adding the oil phase into the aqueous surfactant solution with fixed surfactant concentration. The images on the **second row** are obtained with emulsion droplets prepared by membrane emulsification method in 1.5 wt. % surfactant solution and then diluted with water to the desired surfactant concentration, $d_{ini} \approx 15$ μm. Microscope images of a **(A)** frozen spherical drops; **(B)** frozen, perforated paralellogram flat particle; **(C)** fluid deformed particle which freeze into the same shape; **(D)** flat parallelogram particles – the particle in the centre of the picture is frozen; **(E)** frozen faceted 3D particles; **(F)** frozen long fibres. The total surfactant concentration in the emulsion is (A,B) 0.01 mM and (C,D,E,F) 0.1 mM. The cooling rate is (A,B,C,D) 0.42 K min$^{-1}$ and (E-F) 0.25 K min$^{-1}$. Scale bars: 20 μm.

### 3.2. Surfactant concentration effects for non-ionic surfactants

obtained at fairly high surfactant concentration (1.5 wt. % surfactant concentration, which is 11.75 mM for Tween 40 and 11.56 mM for Tween 60)[4,5,30–32] and the minimum concentration that induces transitions is unknown.

The Brij-class alcohol ethoxylates $C_{18}EO_{10}$ (Brij S10), $C_{18}EO_{20}$ (S20), $C_{16}EO_{10}$ (C10), and $C_{16}EO_{20}$ (Brij58) are well suited to align efficiently at liquid interfaces, to support stabilization of dispersed phases, commonly oil-in-water emulsions. They also induce shape transformations and were previously[5] classified as **Group A** ($C_{18}EO_{10}$, $C_{18}EO_{20}$, $C_{16}EO_{10}$; induce shapes above the melting temperature of the oil) and **Group C** ($C_{16}EO_{20}$, like $C_{16}TAB$, induces shapes below the melting temperature).

We measured the interfacial tension on the hexadecane/water interface as a function of the surfactant concentration at fixed temperature. The values we determined for CMC are 0.022 mM for Tween 40 and 0.011 mM for Tween 60. These are close to values reported in the literature at this temperature (0.033 mM for Tween 40 and 0.0167 mM for Tween 60),[33] though neither of the values is very reliable, and do change for both different suppliers and batches. While the values are not exact, they are by almost 3 orders of magnitude lower than the concentrations at which the self-shaping was previously studied (0.022 mM of Tween 40 is ~500 times more dilute than the initial 1.5 wt. % sample). Just like for the other surfactants, we expect we would see shape changes near these values of CMC. And the CMC we obtained by surface tension are 0.069 mM for Brij S20, 0.0408 mM for Brij 58, 0.126 mM for Brij C10, 0.61 mM for Brij S10.

the surfactant concentration up to 0.03-0.04 mM, the drops start to transform their shape at around 18 °C, but still only reach the initial stages from the evolutionary scheme, including formation of regular polyhedra and other faceted 3D shapes.[4,5,9,10] In the concentration range from 0.03 to 0.1 mM, the transformation and obtained shapes remain the same, see Figure 3C. There is no significant effect from the cooling rate – even at rate as low as ~ 0.05 K min$^{-1}$, no further shapes appear, Figure 3E. At higher surfactant concentrations, ≥ 0.2 mM, we observe formation of flat shapes, which can elongate further into rods if we apply a lower cooling rate or perform the experiment with smaller emulsion droplets, $d_{ini}$ ≤ 15 µm. For this case, it is known that the $C_{16}$ drops stabilized by $C_{16}SorbEO_{20}$ surfactant typically transform their shape to hexagonal or other platelets and relatively stiff rods with diameter ≈ 5 ± 1 µm. Note the variance between the shapes of the droplets on the photographs Figure 3D-E: it is due to the effect of the size of the droplets on the rate of shape transformations, and probably some temperature inhomogeneities.

The results from the second experimental procedure are presented on the second row of Figure 3. In this case, we observe formation of flat platelets even at 0.01 mM Tween 40, see Figure 3B, as opposed to the emulsion droplets prepared by first procedure, see Figure 3A. The observed differences in the

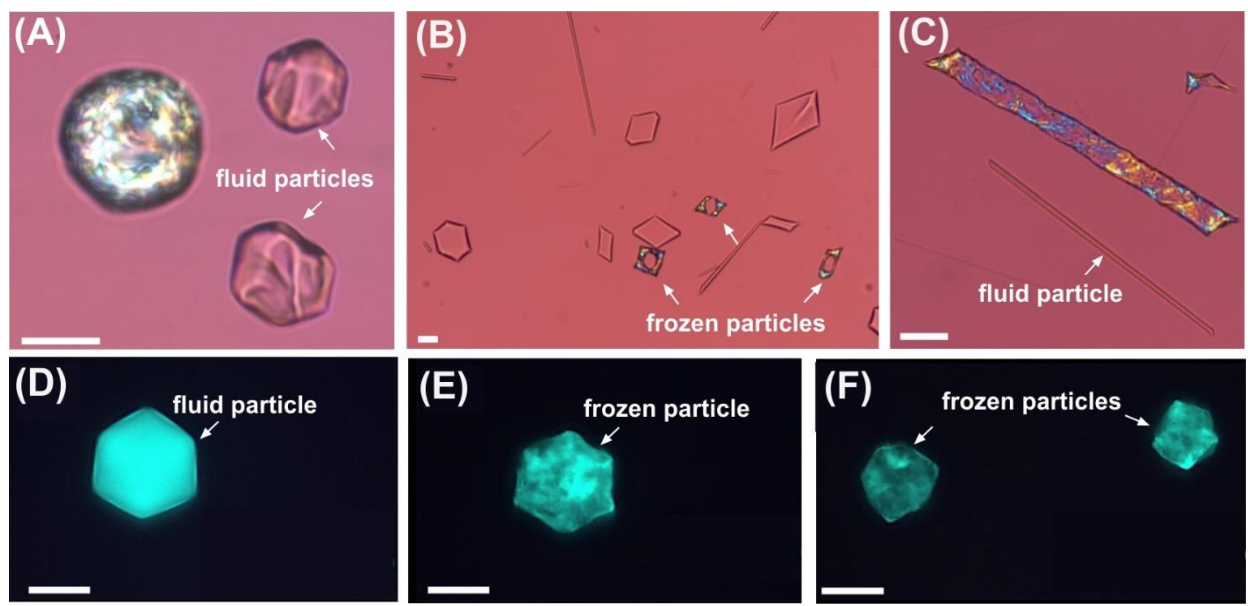

Figure 4: **Drop shape transformations for emulsions of** hexadecane drops in water stabilized by Tween 60. Images on the **first row** are obtained with emulsions prepared by adding the oil phase into the aqueous surfactant solution with fixed surfactant concentration (Procedure 1). The images on the **second row** are obtained with emulsion droplets prepared by membrane emulsification method in 1.5 wt. % surfactant solution and then diluted with water to the desired surfactant concentration (Procedure 2). Droplets $d_{ini}$ ≈ 15 µm. Microscope images of a **(A)** corrugated polyhedra particles; **(B)** flat shapes; **(C)** rods; **(D)** fluid deformed particle; **(E)** frozen deformed hexagon particle; **(F)** frozen deformed particles. The total surfactant concentration in the emulsion is (A) 0.05 mM, (B-C) 1 mM, and (D-F) 2.0x10$^{-4}$ mM. The cooling rate is (A-C) 0.42 K min$^{-1}$, (D-E) 0.2 K min$^{-1}$, and (F) 0.05 K min$^{-1}$. Images in (A-C) are made in transmitted polarized light, while these in (D-F) are in reflected light. Scale bars: 20 µm.

**Experiments with Tween 40 surfactant.**

The procedure for emulsion preparation has a significant effect on the drop shape transformations, see Figure 3. For emulsions prepared by the first experimental procedure, at surfactant concentration of about 0.01 mM, which is about half the value of the measured CMC (0.022-0.033 mM), the $C_{16}$ droplets all freeze into spherical shapes at 12 °C, Figure 3A. If we increase behaviour of the alkane droplets indicate that the denser absorption of surfactant at higher concentration is maintained upon dilution in procedure #2, i.e. no complete equilibration was reached. This could be either due to the presence of a barrier to desorption or (more likely) the difference in kinetics of exchange.

We also analysed the effect of the cooling rate, see Figure 3D and E. It can be seen that depending on the rate of cooling, the droplets are able to reach different stages of the evolutionary scheme, i.e. slower cooling rates ensures more time for the drops to transform their shape and to elongate into rods and ellipsoidal droplets connected by thin fibres. It's shown that the cooling rate does not affect the order of the shape change sequence.

**Experiments with Tween 60 surfactant.**

Additional experiments with Tween 60 ($C_{16}SorbEO_{20}+C_{18}SorbEO_{20}$) allowed us to analyse the effect of the surfactant chain-length. CMC from our interfacial tension isotherm measurements was 0.011 mM, close to literature values for the CMC = 0.017-0.026 mM.

*By the first experimental procedure*, in which we do not perform an additional dilution of the surfactant solution, the minimum surfactant concentration (~ 0.01 mM) necessary to observe drop shape transformations is somewhat below the measured CMC value. At this surfactant concentration, the droplets have enough surfactant molecules on their surfaces to template the plastic crystal and start the self-shaping with formation of corrugated polyhedral, which upon further cooling freeze in this shape. However, the system cannot proceed to the next stages of the self-shaping evolutionary scheme, even at much slower cooling rate of 0.05 K min$^{-1}$, which is known to induce changes more easily.

Further increasing the concentration of the surfactant leads to an increased probability for formation of different shapes. In the concentration range between 0.01 and 0.05 mM, the $C_{16}$ drops are able to transform their shapes only to regular and corrugated polyhedra and irregular 3D shapes, see Figure 4A. The later stages of the evolutionary scheme, such as flat platelet shapes (hexagons, triangles, rhombus), rods and thin fibres, can be reached by increasing the concentration of Tween 60 above 0.1 mM at around 18 °C, see Figure 4B and C. Similarly to the experimental results obtained with Tween 40, the experiments with emulsions prepared by the 2nd experimental procedure yielded lower minimum concentrations necessary to produce self-shaping compared to the 1st procedure. The procedure also affected significantly the final-stage shapes which could be obtained at a given surfactant concentration. Whereas the $C_{16}$ drops stabilized by 1.1×10$^{-3}$ mM Tween 60 (1.5×10$^{-4}$ wt. %) and prepared by the first experimental procedure do not transform and freeze into spherical shape at 12 °C, the drops prepared by the second experimental procedure (with the same final surfactant concentration) transform their shape all the way to flat platelets of triangles and parallelograms. The minimum concentration to induce shapes was much lower with the second procedure, ~ 2×10$^{-4}$ mM (3×10$^{-5}$ wt. %), see Figure 4D, E and F. In this case the drops could reach only the initial stages of the evolutionary scheme (formation of polyhedra or hexagons) upon cooling and eventually freeze in this shape, as shown in Figure 4E.

To analyse whether the final shapes were the result of a fast cooling rate or of surfactant concentration effects, we decreased the cooling rate 4-fold – from 0.2 K min$^{-1}$ to 0.05 K min$^{-1}$, which normally provides sufficient time for all the droplet deformation. Still, similar results were obtained: the hexagon was the furthest shape in the evolution tree formed before droplet freezing, see Figure 4F. Below surfactant concentration of 1.1×10$^{-4}$ mM, i.e. 1.5×10$^{-5}$ wt. %, the process of self-shaping stops completely and all the alkane droplets freeze into a spherical shape. In this case, packing of the molecules is not tight enough, and the plastic phase cannot form, or cannot counteract the corresponding higher surface tension of the droplets at these reduced concentrations, so no self-shaping phenomenon can be observed.

The analysis yields a narrow range for the critical concentrations necessary to induce shapes. For emulsions prepared by the experimental procedure 1, the critical concentration is just below the CMC (0.01mM vs. CMC~ 0.011 mM). When diluting from an emulsion with concentrated surfactant (second procedure), the critical concentration in solution is 50× lower.

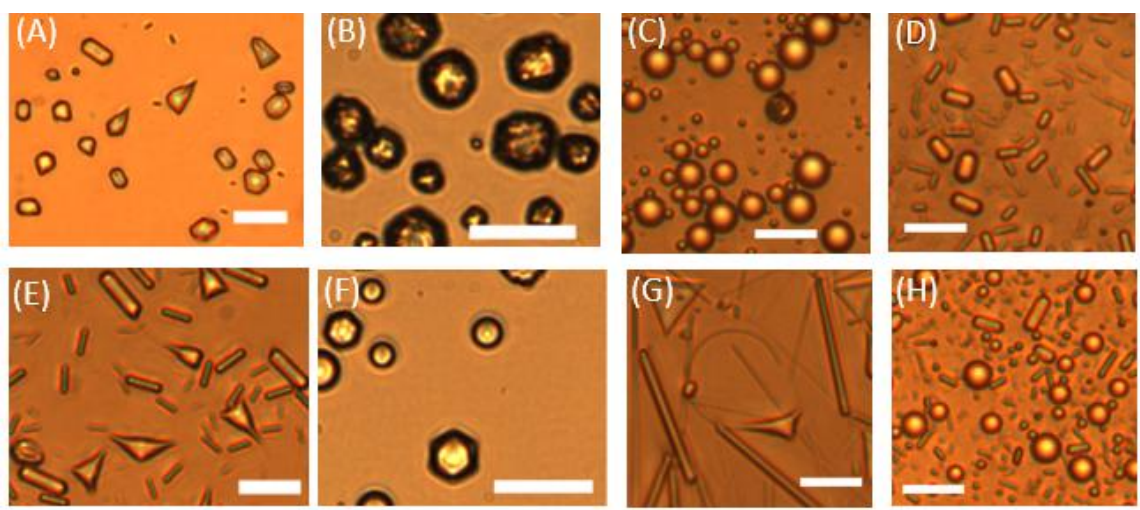

Figure 5: Drop shape transformations for emulsions of hexadecane drops in water stabilized by (A,E) Brij S20, (B,F) Brij 58, (C,G) Brij C10, (D,H) Brij S10. Images on the **first row** are obtained from samples with surfactant concentration just below the CMC. The images on the **second row** are obtained from samples with surfactant concentration just above the CMC. Initial temperature = 20 °C, final temperature = 11 °C and cooling rate is 0.5 - 1 K min$^{-1}$. Scale bar 20 μm.

**Experiments with Brij surfactants.**

Emulsions stabilized by Brij surfactants were prepared by double syringe technique (drops with $d_{ini} \approx$ 5-25 μm). The shape change of droplets in Brij surfactant emulsions were analyzed at four different concentrations, two of which were closely above and below the CMC, whilst the other two were far away from the CMC (measurements for different Brij surfactants in Figure S2).

Our observations confirm that the concentration at which shape changes first appear is close to the CMC, usually around 0.7×CMC the droplets changed shaped at around at 10-12 °C (Figure 5, Table S3). Figure S3-S6 in the supplement material show the droplets freezing/shape changes at different surfactant concentrations. These include the formation of rechargeable droplet swimmers[34] shown in Figure S3 and S7, in the presence of fluorescent dyes. Such minimal systems are of interest both in origin of life studies[35] and in growing bottom-up robotics.[36,37]

self-shaping behaviour of droplets. We chose the non-ionic mixture of the long-tailed Tween 60 ($C_{16}SorbEO_{20}+C_{18}SorbEO_{20}$), which has the ability to self-shape $C_{16}$ drops, and the short-tailed Tween 20 (roughly 40% $C_{12}SorbEO_{20}$, 40% $C_{14}SorbEO_{20}$, 20% $C_{16}SorbEO_{20}$), which does not.[4,5] Keeping the total concentration of surfactant constant (1.5 wt. %), we explored the effect of the balance in the mixture on the ability of droplets to self-shape.

First, a 1:1 wt./wt. surfactant mixture was prepared at a total concentration of 1.5 wt. %. Hexadecane droplets of 15 to 20 μm were cooled at a constant rate of 0.2 K/min, as in the previous experiments. Figure 6A depicts the three different states of surfactant molecules as found in our system – in micelles, a small fraction existing free in solution, and adsorbing and packing at the surface of oil drops. Due to its shorter chain length, Tween 20 cannot effectively template the formation of the alkane rotator phases and disrupts the templating from Tween 60. However, Tween 60 has higher surface activity, so the fraction of Tween 60 at the interface is much larger than the 1:1 ratio in the aqueous solution.

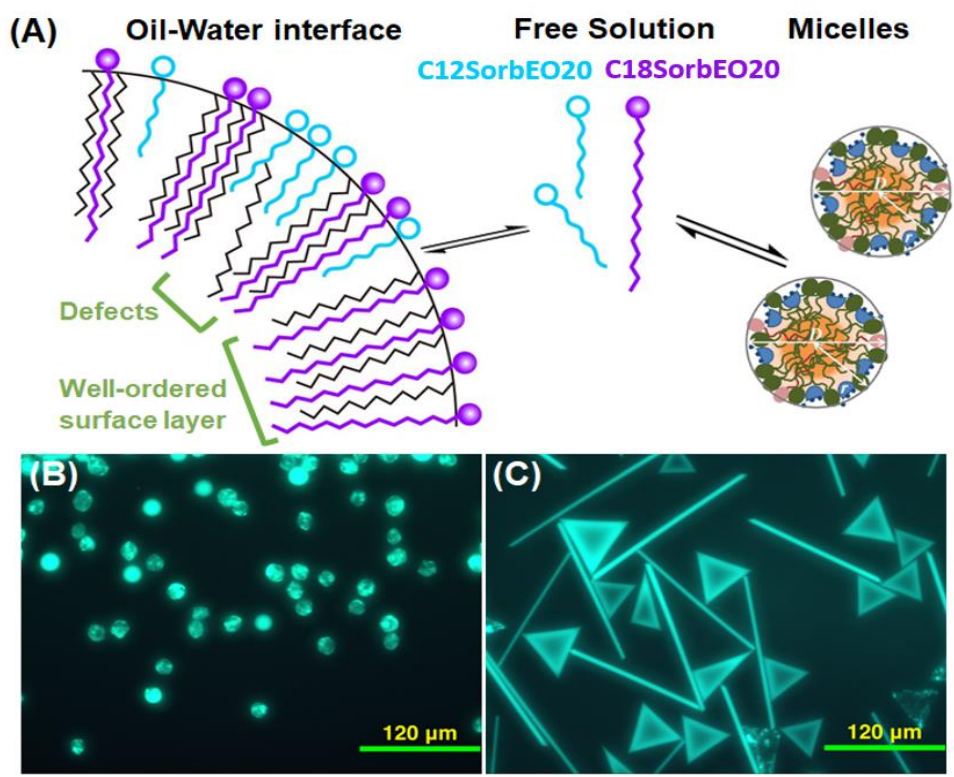

Figure 6: **(A)** Schematic illustration of a mixed surfactant layer containing C18SorbEO20 and C12SorbEO20 at the oil/water interface. C18SorbEO20 may favorably interdigitate with $C_{16}$ alkane molecules to form a well-ordered layer and template the formation of further alkane rotator phase layers beneath it. However, C12SorbEO20 cannot – due to short chain length, this surfactant produces defects in the surface structure. **(B)** 34% C12SorbEO20 occupation. **(C)** 26% C12SorbEO20 occupation.

### 3.3. Competitive adsorption effects on self-shaping in surfactant mixtures

Most commercial surfactants contain significant amounts of impurities, and many industrial formulations include more than one surfactant molecule on purpose. It is therefore important to investigate the effect of the composition of a mixture on the

To determine the concentration of C12SorbEO20 on the drop surface we assume that the components of Tween 20 and Tween 60 mix ideally in the micelles. Under this assumption, accounting for the composition of the two surfactants, and based on measured CMC and adsorption coefficient values, we find that the fraction of Tween 20 in the bulk composition is close to the fraction in the interfacial composition. This means that the large fraction of long chain surfactants in Tween 20 contributes to the effective adsorption on the liquid interface and the apparently low CMC of the Tween20 – see supporting information. We also found that the self-shaping process starts when the bulk fraction of C12SorbEO20 is lower than 20 %, see Table 1.

Table 1

| Wt. ratio in aqueous solution [Tween60]/[Tween20] | % Tween60 | Concentration of $C_{12}SorbEO_{20}$ in mM | % Surface occupation by $C_{12}SorbEO_{20}$ | Appearance of self-shaping |
|---|---|---|---|---|
| 1:73 | 1.35% | 4.8 | 40 | No |
| 1:6.6 | 13.2% | 4.2 | 37 | No |
| 1:3.28 | 23.4% | 3.7 | 34 | No |
| 1:1 | 50% | 2.4 | 26 | Yes |
| 2:1 | 66.7 % | 1.6 | 19 | Yes |

## 4. Conclusions

The experiments above show the minimum concentration of surfactant necessary to induce shape formation is around the surfactants' CMC. We also verify it is not the absolute concentration in solution that is important, but rather the coverage on the water/oil interface. We do this by performing experiments with ionic surfactants in the presence of salts where concentrations that previously could not induce shaping suddenly do, because the presence of salt drastically lowers the CMC, and again we find the critical concentration for inducing shapes is around CMC.

We also perform experiments with a mixed set of non-ionic surfactants with the same heads but one with a tail similar to the alkane oil (Tween 60, $C_{16}SorbEO_{20}+C_{18}SorbEO_{20}$) that induces shapes and one with a much shorter one does not (Tween 20). In experiments we vary the bulk surfactant fraction of $C_{12}SorbEO_{20}$ from 40% to 19% and found that at 26% surface coverage and below of $C_{12}SorbEO_{20}$ we observe self-shaping.

In the context of surface coverage of a surfactant in an experiment with a simple visual readout (as in Figure 6B and C) this approach would allow us to receive accurate information about molecular orientation and packing on the surface (and the formation of rotator phase beneath the surface layer), which was previously only possible to find out by much more expensive, rare, and time-consuming techniques. One could then easily and effectively switch on and off the shape transformations of oil droplets by controlling solution concentrations. Such precise guidance of interfacial layer composition would enable further fundamental experiments into the molecular basis for inducing such phase transitions.

The discovery of minimum requirements needed for shaping droplets has potential for scalable manufacturing of anisotropic structures from monomer droplets.[7] and development of multi-functionality by sequential interpenetration,[38] e.g., super-capacitance,[39] programmable actuation,[38] and self-sensing.[40] Bottom-up, sustainable growth of active particles could be enabled by the minimal surfactant and mixed surfactant criteria established here.

## Author Contributions

We strongly encourage authors to include author contributions and recommend using CRediT for standardised contribution descriptions. Please refer to our general author guidelines for more information about authorship.

J.F. and S.K.S conceived the experiments for minimal surfactants concentration to induce the shapes. In discussions with ST we designed a wider set of experiments including many that were supervised by ST and performed by Zh.V. E.L and E.N. worked on the Brij surfactant's part and analysed the results. TW helped JF in the emulsification and microscope setups. RS and ST theoretically analysed the description of mixed monolayer experiments. J.F. and S.K.S co-wrote the manuscript. J.F., S.S. and Zh.V. created the figures. S.T. and R.S. critically edited the manuscript. All authors contributed to the discussion of results and commented on the manuscript.

## Conflicts of interest

There are no conflicts to declare.

## Acknowledgements

The authors gratefully acknowledge the financial support of grants to Stoyan Smoukov EMATTER (# 280078) and Proof-of-Concept grants ShipShape(#766656) and CoolNanoDrop (#841827) and EPSRC fellowship grant EP/R028915/1. The study falls under the umbrella of European Networks COST CA 17120, COST MP 1305, the Horizon 2020 project "Materials Networking" (ID: 692146-H2020-eu.4.b) and by the Scientific Research Fund of Sofia University (Project No. 80-10-225).

This work is partially supported by the Operational Program "Science and Education for Smart Growth", Bulgaria, grant number BG05M2OP001-1.002-0012.

## Notes and references

‡ Footnotes relating to the main text should appear here. These might include comments relevant not central to the matter under discussion, limited experimental and spectral data, and crystallographic data.

TOC Graphic:

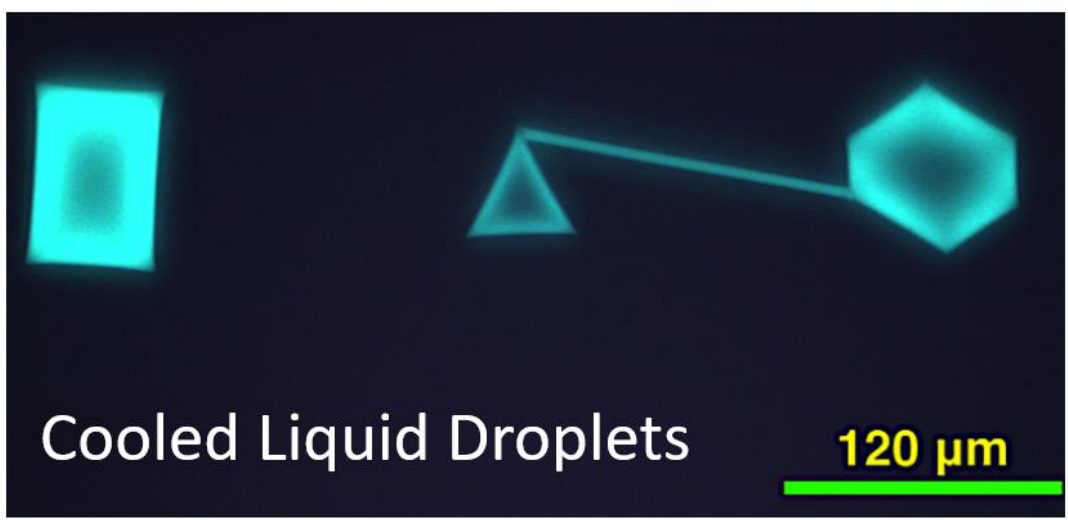

TOC Caption (up to 20 words):
Surfactant above critical micellar concentrations needed for artificial morphogenesis of oil droplets; several surfactant types, salt effects, competitive adsorption investigated.

Supplementary Information for:

# Minimum Surfactant Requirements for Inducing Self-shaping of Oil Droplets and Competitive Adsorption Effects

Jiale Feng[1,5], Zhulieta Valkova[2], E Emily Lin[4], Ehsan Nourafkan[4], Tiesheng Wang[1,3], Slavka Tcholakova[2*], Radomir Slavchov,[4] Stoyan K. Smoukov[1, 2, 4*]

[1] *Active and Intelligent Materials Lab, Department of Materials Science & Metallurgy,*

*University of Cambridge, 27 Charles Babbage Road, Cambridge CB3 OFS, UK*

[2] *Department of Chemical and Pharmaceutical Engineering, Faculty of Chemistry and Pharmacy, Sofia University,*

*1 James Bourchier Ave., 1164 Sofia, Bulgaria*

[3] *School of Mechanical Engineering, Shanghai Jiao Tong University, Shanghai, 200240, China*

[4] *School of Engineering and Materials Science, Queen Mary University of London,*

*Mile End Road, London E1 4NS, UK*

[5] *Cavendish Laboratory, Department of Physics, University of Cambridge, JJ Thomson Avenue, Cambridge CB3*

*OHE, UK*

**Table S1.** Source (Sigma-Aldrich) and properties of the surfactants studied.

| Non-ionic surfactants | Non-ionic surfactant (trade name) | Number of C atoms, n | Number of EO groups, m | HLB | Structural formula |
|---|---|---|---|---|---|
| Polyoxyethylene Sorbitan monoalkylate $C_n SorbEO_{20}$ | Tween 20 | 12, 14, 16 | 20 | 16.7 | 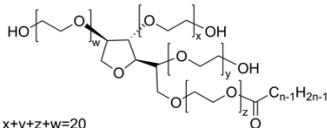 |
| | Tween 40 | 16, 18 | 20 | 15.5 | |
| | Tween 60 | 16, 18 | 20 | 14.9 | |
| **Polyoxyethylene alkyl ethers** $C_nEO_m$ | Brij C10 | 16 | 10 | 12 | $C_nH_{2n+1}-O-(CH_2CH_2O)_m-OH$ |
| | Brij S10 | 18 | 10 | 12 | |
| | Brij 58 | 16 | 20 | 15.7 | |
| | Brij S20 | 18 | 20 | 15.3 | |
| **Ionic surfactant** | **Abbreviation** | **Number of C atoms, n** | **Hydrophilic head group** | **Purity** | **Structural formula** |
| Cetyltrimethyl-ammonium bromide | $C_{16}TAB$ | 16 | $N^+(CH_3)_3Br^-$ | > 99% | $H_3C(H_2C)_{15}-N^+(CH_3)_3 \; Br^-$ |

**Appendix**

Composition of adsorption layer

Tween 20 and Tween 60 mix ideally in the micelles. Under this assumption we can calculate the free concentration of molecules of Tween 20 and Tween 60 in the solution by using eqs. (12) and (9) from Clint[1] together with the experimentally determined values for CMC of Tween 20 and Tween 60:

$$C_1^{monomer} = CMC_1 \frac{\sqrt{\left[\frac{C_{tot}}{CMC_1}-\left(\frac{CMC_2}{CMC_1}-1\right)\right]^2 + 4x_1\frac{C_{tot}}{CMC_1}\left(\frac{CMC_2}{CMC_1}-1\right)} - \left[\frac{C_{tot}}{CMC_1}-\left(\frac{CMC_2}{CMC_1}-1\right)\right]}{2\left(\frac{CMC_2}{CMC_1}-1\right)}, \quad (1a)$$

$$C_2^{monomer} = \left(1 - \frac{C_1^{monomer}}{CMC_1}\right)CMC_2 \quad (1b)$$

Here, $C_1^{monomer}$ and $C_2^{monomer}$ are the free monomer concentration of Tween 20 and Tween 60 in the aqueous solution, $x_1$ is the molar fraction of Tween 20 and $C_{tot}$ is the total surfactant concentration. Assuming that the molecular masses of Tween 20 and Tween 60 are 1228 g/mol and 1298 g/mol, we can calculate the free concentration of monomers in the solution. In order to estimate the ratio between Tween 20 and Tween 60 on the oil-water interface we used the expression for competitive adsorption on the interface assuming a Langmuir mixed adsorption isotherm[2]:

$$y_2 = \frac{\Gamma_2}{\Gamma_1 + \Gamma_2} = \frac{K_{A2}C_2^{monomer}}{K_{A1}C_1^{monomer} + K_{A2}C_2^{monomer}} \quad (2)$$

where $y_2$ is the mole fraction of 2$^{nd}$ component (Tween 60) on the interface, $\Gamma$ are the interface adsorptions of the respective components and $K_A$ are the adsorption constants on the drop-water interface of the surfactant molecules, and $C_1^{monomer}$ and $C_2^{monomer}$ are the free monomer concentration of Tween 20 and Tween 60.

From interfacial tension isotherms we determined that CMC for Tween 20 is 0.0167 mM, whereas CMC for Tween 60 is 0.0108 mM. From the best fit of the experimental data around CMC we determined that $\Gamma_{CMC}$ are 4.3 µmol/m$^2$ and 4.4 µmol/m$^2$ for Tween 20 and Tween 60, respectively. From the measured interfacial tensions below CMC by using the Langmuir adsorption isotherm we determined that the adsorption constants are 2200 m$^3$/mol and 1900 m$^3$/mol for Tween 20 and Tween 60, respectively. Using these values we determine the composition of adsorption layers given in Table S2.

The obtained results showed that the composition of the aqueous phase is close to the composition of the interface which is related to the fact that the determined CMC and adsorption constant of Tween 60 and Tween 20 do not differ significantly, which means that there is a significant fraction of long chain surfactants in Tween 20 which are able to adsorb on the solution surface and to decrease Tween-20's CMC.

Table S2

| Wt. ratio in aqueous solution [Tween60]/[Tween20] % Tween60 | Free monomer concentration of Tween 60 in µM | Surface occupation by Tween 60 (%) From Eqn. 2 | Appearance of self-shaping |
|---|---|---|---|
| 1:73 | 1.35% | 0.14 | 0.72 | No |
| 1:6.6 | 13.2% | 1.3 | 7.4 | No |
| 1:3.28 | 23.4% | 2.4 | 14 | No |
| 1:1 | 50% | 5.2 | 35 | Yes |
| 2:1 | 66.7 % | 7.1 | 51 | Yes |

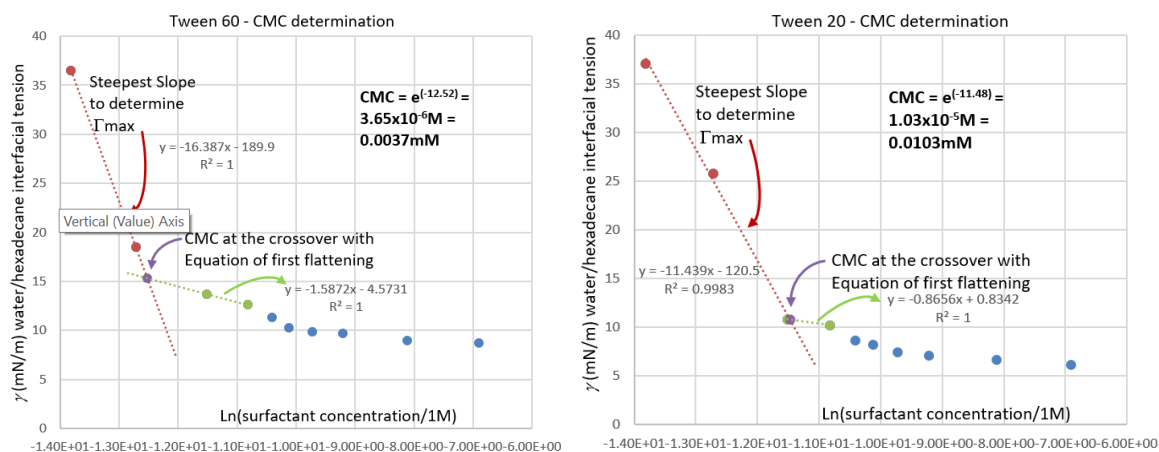

Figure S1. Adsorption isotherms for Tween 20 and Tween 60 at hexadecane-water interface at 20 °C.

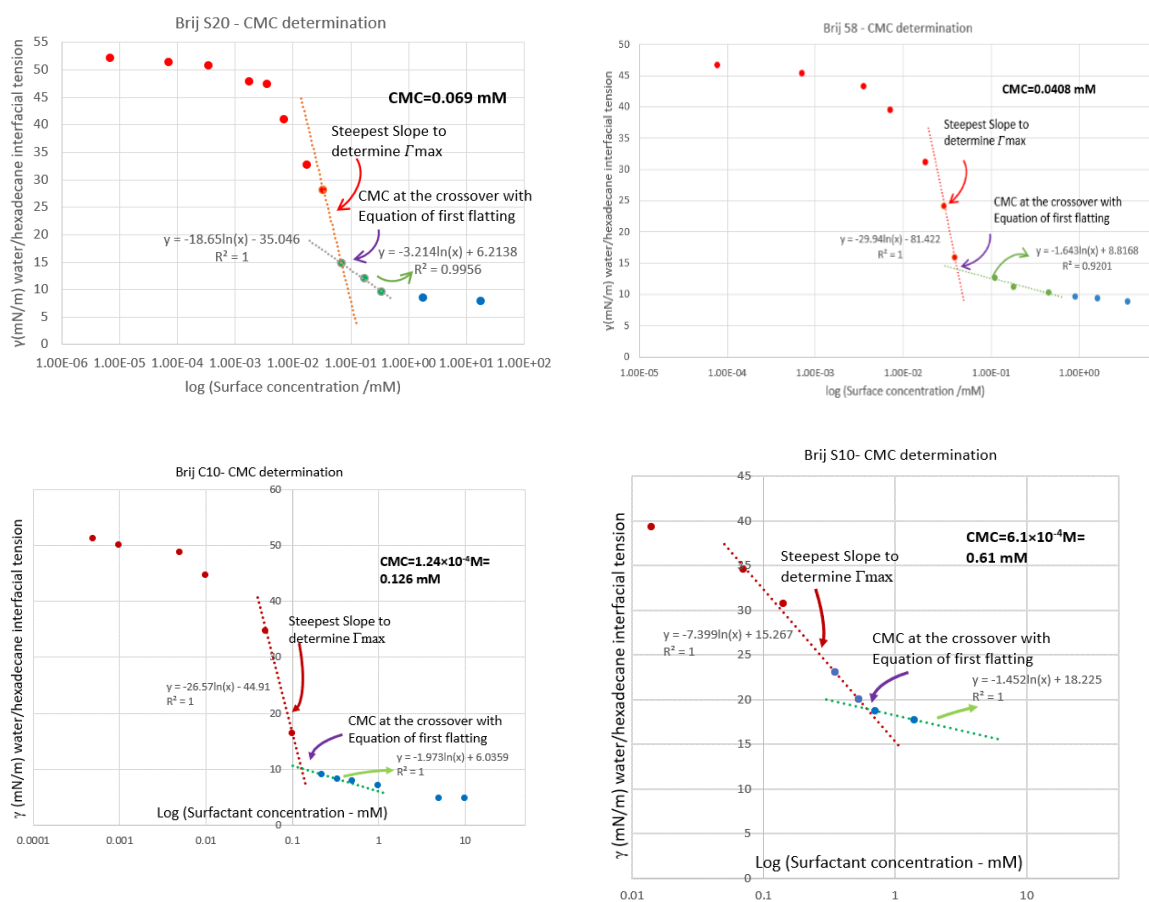

Figure S2. Adsorption isotherms for Brij S20, Brij 58, Brij C10 and Brij S10 at hexadecane-water interface at 20 °C.

Table S3

| Surfactant | Concentration (mM) | Multiple of CMC | Appearance of self-shaping |
|---|---|---|---|
| Brij S20 | 0.21 | 3 | Yes |
| | 0.13 | 1.9 | Yes |
| | 0.034 | 0.49 | Yes |
| | 0.0093 | 0.13 | No |
| Brij 58 | 0.41 | 10 | Yes |
| | 0.3 | 6.5 | Yes |
| | 0.06 | 1.3 | Yes |
| | 0.03 | 0.65 | No |
| Brij S10 | 0.7 | 1.15 | Yes |
| | 0.5 | 0.86 | Partially self-shaping |
| | 0.35 | 0.57 | Partially self-shaping |
| | 0.14 | 0.23 | No |
| Brij C10 | 0.5 | 3.96 | Yes |
| | 0.3 | 2.6 | Yes |
| | 0.2 | 1.7 | Yes |
| | 0.1 | 0.7 | No |

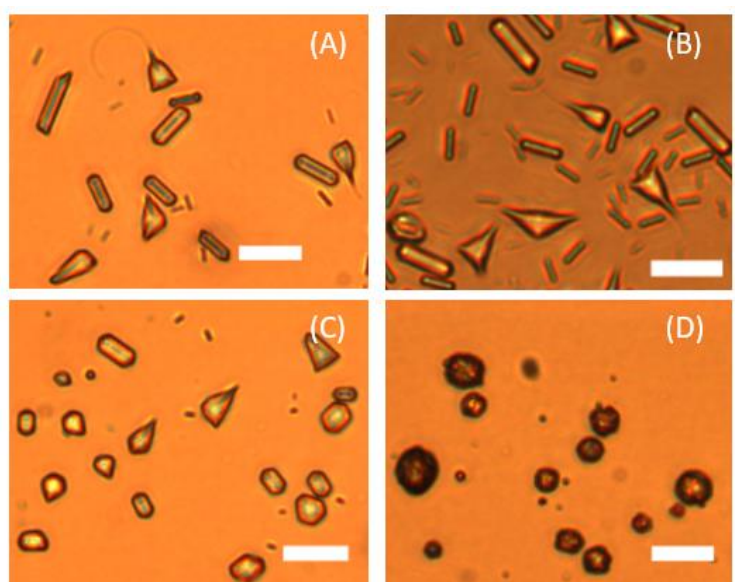

Figure S3: **Drop shape transformations for emulsions of** hexadecane drops in water stabilized by Brij S20. Cooling rate is 0.5 K min$^{-1}$. The total surfactant concentration in the emulsion is (A) 0.21 mM, (B) 0.13 mM, (C) 0.034 mM, (D) 0.0093 mM Scale bars: 20 μm.

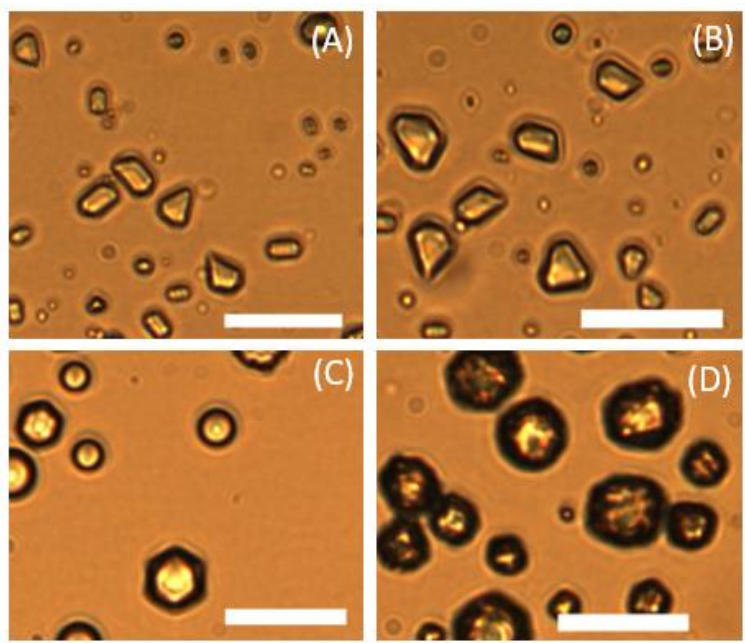

Figure S4: **Drop shape transformations for emulsions of** hexadecane drops in water stabilized by Brij 58. Cooling rate is 0.5 K min$^{-1}$. The total surfactant concentration in the emulsion is (A) 0.41 mM, (B) 0.3 mM, (C) 0.06 mM, (D) 0.03 mM Scale bars: 20 μm.

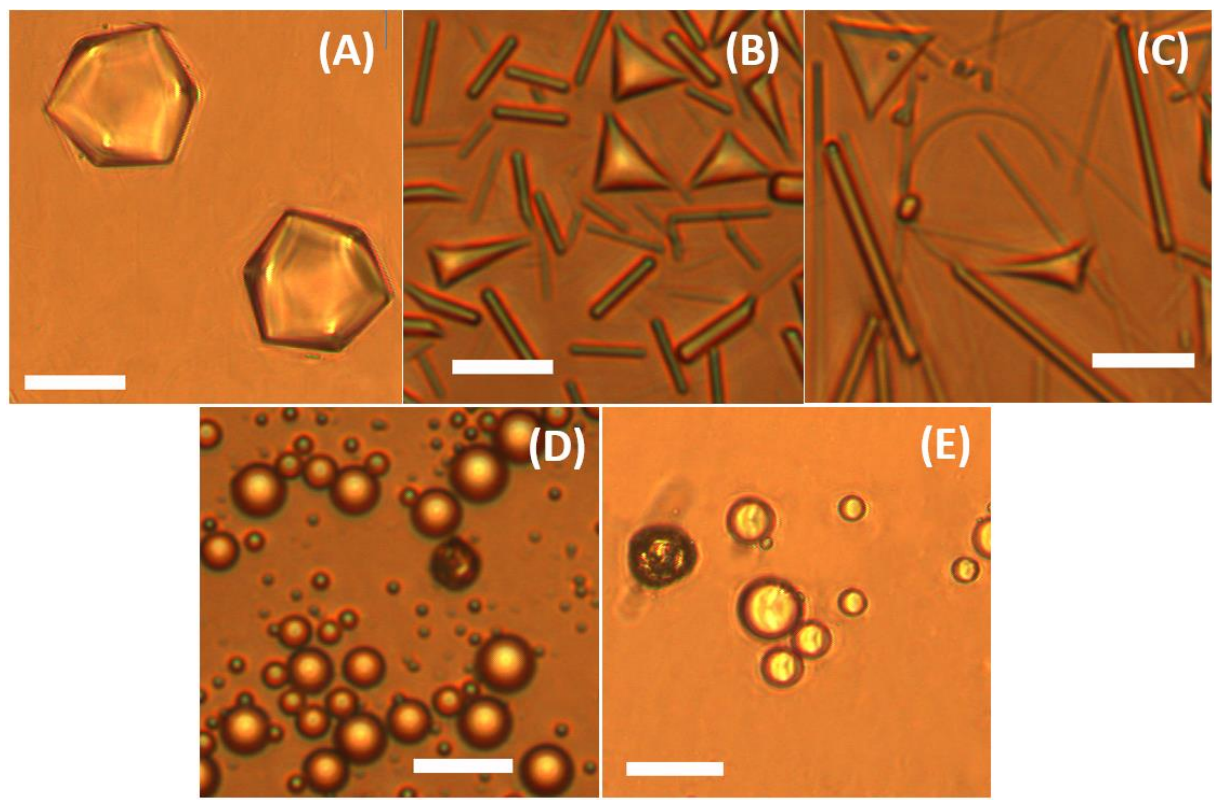

Figure S5: **Drop shape transformations for emulsions of** hexadecane drops in water stabilized by Brij C10. Images on the **first row** are obtained with emulsions prepared by the double syringe emulsification technique. Droplets $d_{ini} \approx 15$ µm and the cooling rate is 0.5 K min$^{-1}$. The total surfactant concentration in the emulsion is (A) 0.5 mM, (B) 0.3 mM, (C) 0.2 mM, (D) 0.1 mM and (E) 0.05 mM. Scale bars: 20 µm.

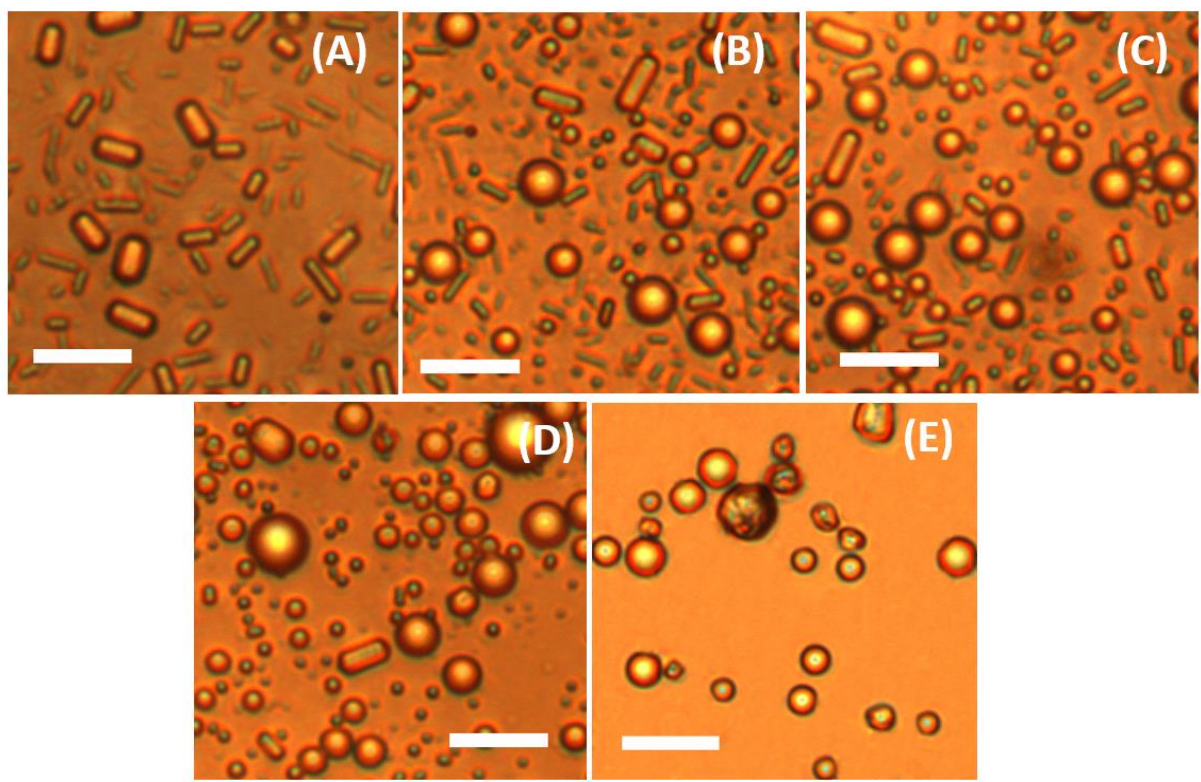

Figure S6: **Drop shape transformations for emulsions of** hexadecane drops in water stabilized by Brij S10. Images on the **first row** are obtained with emulsions prepared by the double syringe emulsification technique. Droplets $d_{ini} \approx 15$ µm and the cooling rate is 0.5 K min$^{-1}$. The total surfactant concentration in the emulsion is (A) 0.7 mM, (B) 0.5 mM, (C) 0.35 mM, (D) 0.14 mM and (E) 0.07 mM. Scale bars: 20 µm.

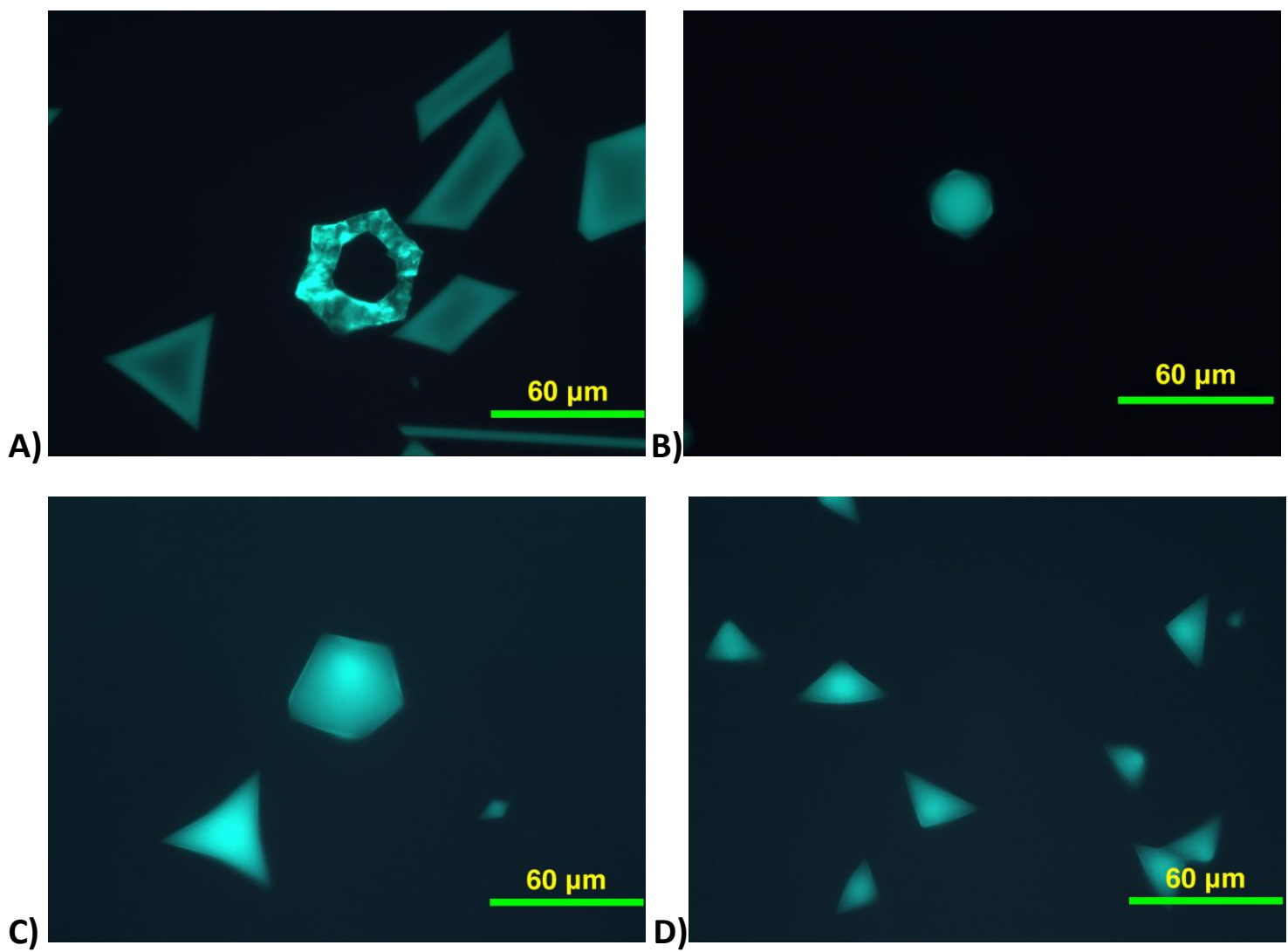

Figure S7. Fluorescence micrographs of hexadecane droplets containing 0.001% fluorescent dye (Solvent Green 5), in aqueous solutions with various non-ionic surfactants. All solutions contained 1.5 wt.% of surfactant in the water phase and were cooled at 0.2 K/min. (A) Tween 40. Frozen hexagonal droplet surrounded by liquid droplets of other shapes. B) Brij 58. C, D) Brij S20. C) shows initial deformations at temp ~ 15 °C, while D) shows later deformation to swimming shapes, as documented in our recent paper,[3] and Figure S3, at temp ~ 14 °C.